\documentstyle[graphicx,preprint,tighten,aps]{revtex}

\begin{document}
\title{Statistical analysis of light fragment production from medium
energy proton-induced reactions} 
\author{S. Furihata\thanks{Fax : +1-604-222-1074, email: furihata@triumf.ca}} 
\address{TRIUMF, Vancouver, British Columbia, Canada V6T 2A3} 
\address{Mitsubishi Research Institute Inc., 2-3-6 Otemachi,
Chiyoda-ku, Tokyo 100-8141, Japan}
\date{\today}
\maketitle
\begin{abstract}
The light fragment production reactions for 10 MeV to 3 GeV
protons incident on $^{16}$O, $^{27}$Al, $^{\rm nat}$Fe, $^{93}$Nb, and
$^{\rm nat}$Ag are analyzed by a combination of an
intranuclear cascade model and a generalized evaporation model
 which includes light nuclei up to Mg as ejectiles. 
It is concluded that evaporation 
is the dominant process by which particles lighter than or equal
to Be are produced from targets heavier than O.
\end{abstract}
\narrowtext

\section{Introduction}

Fragment and residual nuclei production has attracted many
people's interests, not only nuclear physicists but also astrophysicists
and nuclear engineers. 
The production of intermediate mass fragments from high energy
 proton-nucleus reactions or nucleus-nucleus reactions has been a hot
 topic in nuclear physics for a decade \cite{ref98,ref54}.
Astrophysics and cosmic ray physics have been interested in residual
 nuclei production in order to calculate the production of cosmogenic
 nuclides in extraterrestrial matter by solar and galactic cosmic rays.
Recently, nuclear engineering has needed the particle production cross
sections for the 
 development of accelerator-based systems for transmutation of
radioactive  nuclear waste. 
From the radiation safety aspect, it has also become more important to
 estimate the amount of radioactivity produced from various targets,
 as new applications of high energy proton accelerators, such as
 spallation neutron sources and the production of beams of unstable
nuclei are  being developed. 

In 1997, the Organization for Economic Cooperation and Development
(OECD) Nuclear Energy Agency (NEA) conducted benchmark 
calculations on activation 
yields to determine the predictive power of current nuclear reaction
models and codes \cite{NEA}. 
The results calculated using many different codes which are based on a 
combination of different models, such as the intranuclear cascade
model (INC), the exciton model, the evaporation-fission model, 
the quantum molecular dynamics model and the statistical 
multifragmentation model, were compared with experimental data.
It became clear that most of the computer codes did not reproduce
light fragment production reactions, such as Fe(p,X)$^7$Be,
especially at low proton-incident energy. 
They considered that an adequate description of the Fermi break-up
model \cite{ref59} and a fragmentation model was urgently needed.

The evaporation model has been very successful in describing 
residual nuclei production from hot nuclei. 
In many codes, not only those codes that were used in the benchmark
calculation by OECD/NEA, but also the codes that are widely used for
shielding calculations, such as the LAHET code \cite{LAHET}, the evaporation
model is used to describe 
the de-excitation of thermalized nuclei. 
Despite its success, the model has not been used to describe light
fragment emission, except for a few studies concerning break-up of
highly excited nuclei \cite{ref98,ref62}.   

In this study, we propose a generalized evaporation model (GEVAP) 
for Monte Carlo simulation, based on the Weisskopf-Ewing model
\cite{wei2,wei1}.  
Nucleons and helium nuclei are the dominant particles emitted from an
excited nucleus.
Therefore, only these particles are treated as ejectiles in the
Dostrovsky's evaporation models\cite{evap} implemented in the LAHET
code \cite{LAHET}.
On the other hand, some studies \cite{ref98,ref62} consider
light nuclei heavier than $\alpha$ particles as ejectiles since there 
is no reason that those particles can not be emitted from excited
nuclei via evaporation process.
In our generalized evaporation model, 66 nuclides up to Mg are
included as ejectiles, not only in their ground states but also in
their excited states. 
Besides, we use the accurate level density function for the total decay
width calculation instead of an approximate form of level density
function  which is used in the Dostrovsky's evaporation models
\cite{evap}. 

Light fragments produced from proton-induced reactions are analyzed by 
the combination of the INC model implemented in the LAHET code
\cite{LAHET} and the generalized evaporation model (GEVAP). 
In order to estimate light particle production from a nucleon-nucleus 
reaction by the generalized evaporation model, we have to assume the
ensemble of hot nuclei which are produced after the initial
non-equilibrium stage.    
Since the excitation energy, the mass, and the charge
of the hot residual nuclei produced from high energy reaction are
widely distributed, we can not use the simple assumption that a single
excited nucleus represents the ensemble of hot thermalized nuclei. 
However, the INC model can provide an ensemble of
residual nuclei with broad distribution in excitation energy, nuclear
mass and charge. 
The LAHET code employs the Bertini intranuclear cascade model
\cite{Bertini} for a non-equilibrium stage of nuclear reaction,
and the Fermi break-up model \cite{ref59} and the evaporation model
proposed by Dostrovsky {\it et al.} \cite{evap} for a thermalized 
stage. 
Mass $A_i$, charge $Z_i$, excitation energy $E$, recoil energy,
and the direction of recoil motion are extracted from the
INC calculation done by the LAHET code.
Then the de-excitation process of the hot nucleus with these quantities are
calculated by GEVAP, instead of by the Fermi break-up model and the
evaporation model employed in the LAHET code.  
In the following, we call this calculation procedure `INC/GEVAP'.   

We focus mainly on $^7$Be produced by proton-induced reactions in the
energy range from 10 MeV to 3 GeV,
because $^7$Be is the most intensively measured light fragment
produced from various targets, and many
experimental data are available for comparison.
We compare the INC/GEVAP results with experimental data as well as 
the results calculated by using LAHET, to make the effect of using different 
de-excitation models clear.   

\section{The generalized evaporation model}\label{s-gevap}
Let us consider that a parent nucleus $i$ with an excited energy
$E$[MeV], a mass number $A_i$, and a charge number $Z_i$ emits a particle $j$
in its ground state with $A_j$ and $Z_j$, and becomes a daughter
nucleus $d$ with $A_d$ and $Z_d$.  
According to the Weisskopf's formulation \cite{wei2}, the decay probability
$P_j$ with total kinetic energy in the center-of-mass system between
$\epsilon$ and $\epsilon$ + d$\epsilon$ is expressed as
\begin{equation}
P_j(\epsilon)d\epsilon=g_j \sigma_{inv}(\epsilon) 
\displaystyle{\frac{\rho_d(E-Q-\epsilon)}{\rho_i(E)}}
 \epsilon d\epsilon ,
\label{e-gam1}
\end{equation}
where $\sigma_{inv}$ is the cross section for the inverse reaction, 
$\rho_i$ and $\rho_d$ are level densities [MeV$^{-1}$] of the parent
and the daughter nucleus, respectively.  
With the spin $S_j$ and the mass $m_j$ of the emitted particle $j$,
$g_j$ is expressed as $g_j=(2S_j +1) m_j/\pi^2 \hbar^2$.
In this study we use the Audi-Wapstra mass table \cite{AWmass} to
calculate the Q-values $Q$ for emission of particle $j$.

The cross section for the inverse reaction $\sigma_{inv}$ is expressed
as \cite{evap}
\begin{equation}
\sigma_{inv}(\epsilon) = \left\{\begin{array}{ll}
\sigma_g c_n \left( 1 + b/\epsilon\right) & {\rm for\  neutrons}\\
\sigma_g c_j \left( 1 -V/\epsilon\right) & {\rm for\ charged\ particles}\\
				\end{array}
\right.
\equiv \sigma_g \alpha \left( 1 + \frac\beta\epsilon\right), \label{e-siginv} 
\end{equation}
where $\sigma_g=\pi {R_b}^2$ [fm$^2$] is the geometric cross section,
and $V=Z_jZ_d e^2/R_c$ is the Coulomb barrier.

In this study, we use the parameter set determined by Dostrovsky
{\it et al.} \cite{evap} and Matsuse {\it et al} \cite{ref124}.
Dostrovsky {\it et al.} \cite{evap} determined $c_n$, $c_j$, $b$,
$R_b$, and $R_c$  for n, p, d, t, $^3$He, and $\alpha$ emission by
fitting the expression to the theoretical calculation done by Shapiro
\cite{shapiro} and Blatt and Weisskopf \cite{blatt}, so that the
effect of overlapping wave functions was taken into account. 
These parameters are used in the Dostrovsky's evaporation 
model\cite{evap} implemented in LAHET\cite{LAHET}.
Meanwhile, Matsuse {\it et al.} determined the critical distance 
($R_b$ and $R_c$, with $c_j=1$) by fitting Eq. (\ref{e-siginv}) to
experimental fusion cross sections for heavy ion reactions.
We use the Dostrovsky's parameters for n, p, d, t, $^3$He, and
$\alpha$ emission and the Matsuse's parameters for other particles.
In the following we call these parameters ``the precise parameter set''.
Besides the calculation with the precise parameter set, we use the
simple parameter set, given by
$c_n=c_j=1$, $b=0$ and $R_b=R_c=r_0(A_j^{1/3}+A_d^{1/3})$ [fm]  for
the inverse cross section.
In the calculation with the simple parameter set, values of 
$r_0=1.2$, $1.5$, and $2.0$ are tried to test the stability of our model.

The total decay width $\Gamma_j$ can be calculated by integrating
Eq.\ (\ref{e-gam1}) with respect to the total kinetic energy $\epsilon$ 
from the Coulomb barrier $V$  up to the maximum possible value
$(E-Q)$.   
By using Eq.\ (\ref{e-siginv}) for $\sigma_{inv}$, 
the total decay width for the particle emission is expressed as
\begin{equation}
\Gamma_j= \displaystyle{\frac{g_j \sigma_g \alpha}{\rho_i(E)}}
\displaystyle{\int_{V}^{E-Q}}\epsilon
\left(1 + \frac\beta\epsilon \right) \rho_d(E-Q-\epsilon)  d\epsilon .
\label{e-gam2}
\end{equation}

According to the Fermi-gas model,
the total level density $\rho(E)$ of a nucleus summed over all the
possible states with the angular momenta is given by the expression
\cite{Gilcam}  
\begin{equation}
\rho(E) =
 \frac\pi{12} \frac{e^{2\sqrt{a(E-\delta)}}}{a^{1/4}(E-\delta)^{5/4}} 
\quad \mbox{ for }  E\geq E_x \label{rho1},
\end{equation}
where $a=A_d/8$ [MeV$^{-1}$] is the level density parameter, and
$\delta$[MeV] is the pairing energy of the daughter nucleus
evaluated by Cook {\it et al.} \cite{COOK}.
For those values not evaluated by Cook {\it et al.},
$\delta$ obtained by Gilbert and Cameron \cite{Gilcam} are used.
$E_x$ is determined by Gilbert and Cameron \cite{Gilcam} as 
$E_x=U_x+\delta$ where $U_x=2.5 + 150/A_d$.
In the calculation with the precise parameter set, we use 
the Gilbert-Cameron-Cook-Ignatyuk (GCCI) level density parameter
\cite{LAHET}, 
in which the pairing corrections and the energy dependence of the level
density parameter are taken into account,
instead of the simple expression $a=A_d/8$.
The GCCI level density parameter is employed in the
LAHET code\cite{LAHET}. 

When $E$ is below $E_x$, instead of Eq.\ (\ref{rho1}) the following
formula gives a good fit to the experimental level densities
\cite{Gilcam}: 
\begin{equation}
\rho(E)= \frac1T e^{(E-E_0)/T} \text{ for }  E<E_x, \label{rho2}
\end{equation}
where $T$ is the nuclear temperature given by $1/T= \sqrt{a/U_x}-1.5/U_x$.
To connect Eq.\ (\ref{rho1}) and Eq.\ (\ref{rho2}) smoothly, $E_0$ is
defined as 
$E_0=E_x-T(\log T - 0.25 \log a -1.25 \log U_x+ 2 \sqrt{aU_x})$.

We use the expressions Eq.\ (\ref{rho1}) and Eq.\ (\ref{rho2}) to
calculate the total decay width.
The simple form $\rho \propto \exp(2\sqrt{a(E-\delta)})$, which
is used in the Dostrovsky's evaporation models \cite{evap}, 
is a good approximation when the residual excitation energy is high,
however, it is not applicable for residual nuclei with small mass and
low excitation energy. 

When $E-Q-V$ is below $E_x$, 
Eq.\ (\ref{e-gam2}) can be solved analytically, by substituting
Eq.\ (\ref{rho2}) into Eq.\ (\ref{e-gam2}).
\begin{equation}
\Gamma_j=\frac{\pi  g_j \sigma_g\alpha}{12\rho_i(E)} \{I_1(t,t)+(\beta+V)
I_0(t)\} \text{ for } E-Q-V < E_x, \label{e-gamma}
\end{equation}
where $I_0(t)$ and $I_1(t,t_x)$ are expressed as:
\begin{eqnarray}
I_0(t)&=& e^{-E_0/T}(e^t-1),\nonumber\\
I_1(t,t_x)&=&e^{-E_0/T}T \{(t-t_x+1)e^{t_x}-t -
1)\},\nonumber 
\end{eqnarray}
where $t=(E-Q-V)/T$ and $t_x=E_x/T$.
When $E-Q-V$ is greater than $E_x$, the integral of
Eq.\ (\ref{e-gam2}) can not be solved analytically because of the
denominator in Eq.\ (\ref{rho1}).
However, it is expressed approximately as
\widetext
\begin{equation}
\Gamma_j=\frac{\pi  g_j \sigma_g\alpha}{12\rho_i(E)} 
 \left[I_1(t,t_x)+ I_3(s,s_x)e^s +(\beta+V)
 \left\{I_0(t_x)+I_2(s,s_x)e^s\right\}\right] 
 \text{ for }  E-Q-V \geq E_x. \label{e-gamma2} 
\end{equation}
where $I_2(s,s_x)$ and $I_3(s,s_x)$ are given by:
\begin{eqnarray*}
I_2(s,s_x)&=& 2\sqrt{2}\left\{
s^{-3/2}+1.5s^{-5/2}+3.75s^{-7/2}
-(s_x^{-3/2}+1.5s_x^{-5/2}+3.75s_x^{-7/2})e^{s_x-s}
\right\},\\
I_3(s,s_x)&=& (\sqrt{2}a)^{-1}\left[
2s^{-1/2}+4s^{-3/2}+13.5s^{-5/2}+60.0s^{-7/2}+325.125s^{-9/2}
\right. -\left\{ (s^2-s_x^2)s_x^{-3/2} \right.\\
&+&(1.5s^2+0.5s_x^2)s_x^{-5/2}+(3.75s^2+0.25s_x^2)s_x^{-7/2} 
+(12.875s^2 +0.625s_x^2)s_x^{-9/2} \\
&+&(59.0625s^2+0.9375s_x^2)s_x^{-11/2}+\left.\left.(324.8s^2+
3.28s_x^2)s_x^{-13/2}\right\}e^{s_x-s}\right], 
\end{eqnarray*}
\narrowtext
with $s=2\sqrt{a(E-Q-V-\delta)}$ and $s_x=2\sqrt{a(E_x-\delta)}$.

In the present  Monte Carlo simulation, ejectile $j$ is selected according 
to the probability distribution calculated as $p_j=\Gamma_j/\sum_j
\Gamma_j$, where $\Gamma_j$ is given by Eqs. (\ref{e-gamma}) or
(\ref{e-gamma2}). 
The total kinetic energy $\epsilon$ of the emitted particle $j$ and
the daughter nucleus is chosen according to the probability
distribution given by Eq. (\ref{e-gam1}).
The angular distribution of the motion is randomly selected to be
isotropic in the center-of-mass system. 
The excitation energy of the daughter nucleus $E_d$ is calculated as
$E_d=E-Q-\epsilon$.

In this study, we consider 66 nuclides as ejectiles, not only in their
ground states but also in their excited states. 
It is important to include excited states in the particles emitted via
the evaporation process, because it greatly enhances the yield of
heavy particle emission \cite{ref62}. 
The selected ejectiles satisfy the following criteria: 
(1) isotopes with $Z_j\leq 12$; 
(2) naturally existing isotopes or isotopes near the stability line; 
(3) isotopes with half-life longer than 1 ms.
The selected ejectiles are listed in Table \ref{t-emittor}.

If the mean lifetime of a resonance is longer than the decay width of
the resonance emission, such a resonance can survive during the
evaporation process.    
The excited state is included if its half lifetime $T_{1/2}$ [sec]
satisfies the following condition: 
\begin{equation}
\frac{T_{1/2}}{\ln 2}> \displaystyle{\frac{\hbar }{\Gamma_j^*}},
\label{e-condition}
\end{equation}
where $\Gamma_j^*$ is the decay width of the resonance emission.
$\Gamma_j^*$ can be calculated in the same manner as for a ground state
particle emission. 
The Q-value for the resonance emission is expressed as
$Q^*=Q+E_j^*$, where $E_j^*$ is the excitation energy of the resonance.
The spin state of the resonance $S_j^*$ is used in the calculation of
$g_j$, instead of the spin of the ground state $S_j$.
We use the ground state mass $m_j$ for excited states because
the difference between the masses is negligible. 

Instead of treating a resonance as an independent particle,
we simply enhance the decay width of the ground state particle
emission. 
We redefine the decay width $\Gamma_j$ as 
\begin{equation}
\Gamma_j=\Gamma_j^0+\displaystyle{\sum_{n} \Gamma_j^n} ,
\end{equation}
where $\Gamma_j^0$ is the decay width of the ground state particle
$j$ emission, and $\Gamma_j^n$ is that of the $n$th excited state of the
particle $j$ emission which satisfies Eq.\ (\ref{e-condition}).

The total kinetic energy distribution of the excited particle emission is
assumed to be the same as that of the ground state particle emission.
$S_j^*$, $E_j^*$, and $T_{1/2}$ used in this study are extracted from the
Evaluated Nuclear Structure Data File (ENSDF) database maintained by
the National Nuclear Data Center\cite{NNDC}.  


\section{Comparison with experimental data}\label{s-res}

\begin{figure}
\begin{center}
\begin{tabular}{ll}
 \rotatebox{90}{  
\includegraphics[%
         height=7cm]{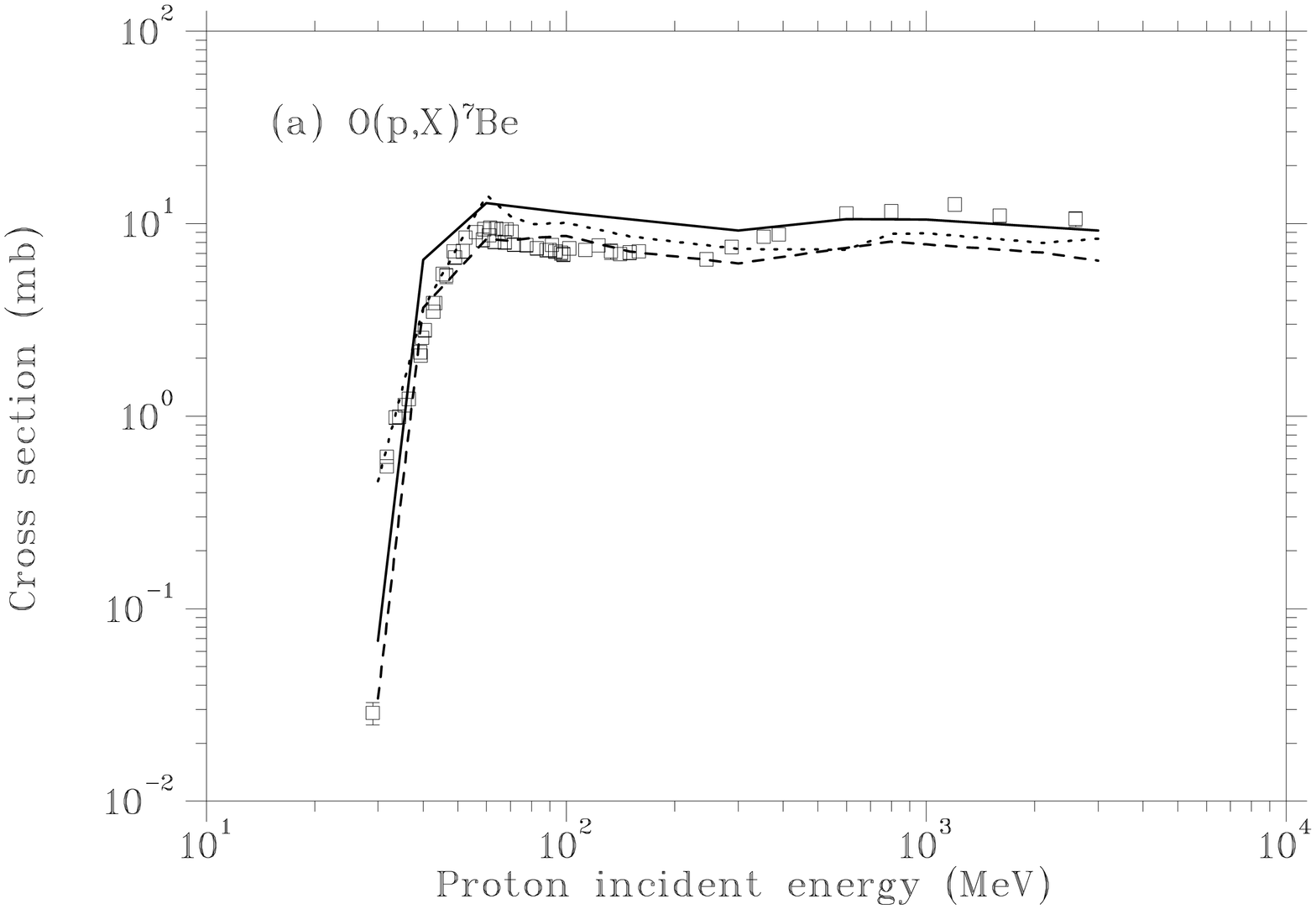}}
&
 \rotatebox{90}{  
\includegraphics[%
         height=7cm]{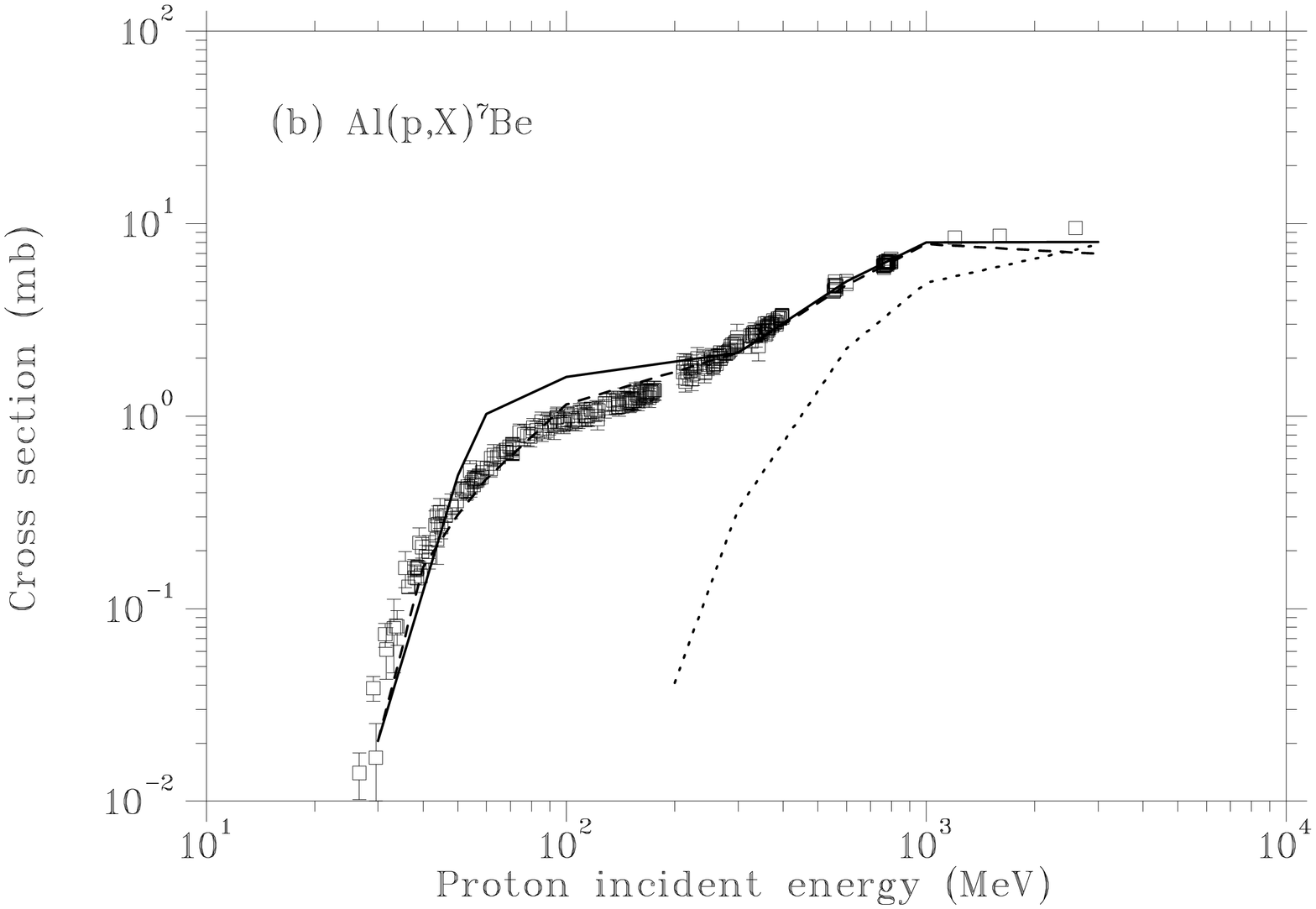}
}\\
 \rotatebox{90}{  
\includegraphics[%
         height=7cm]{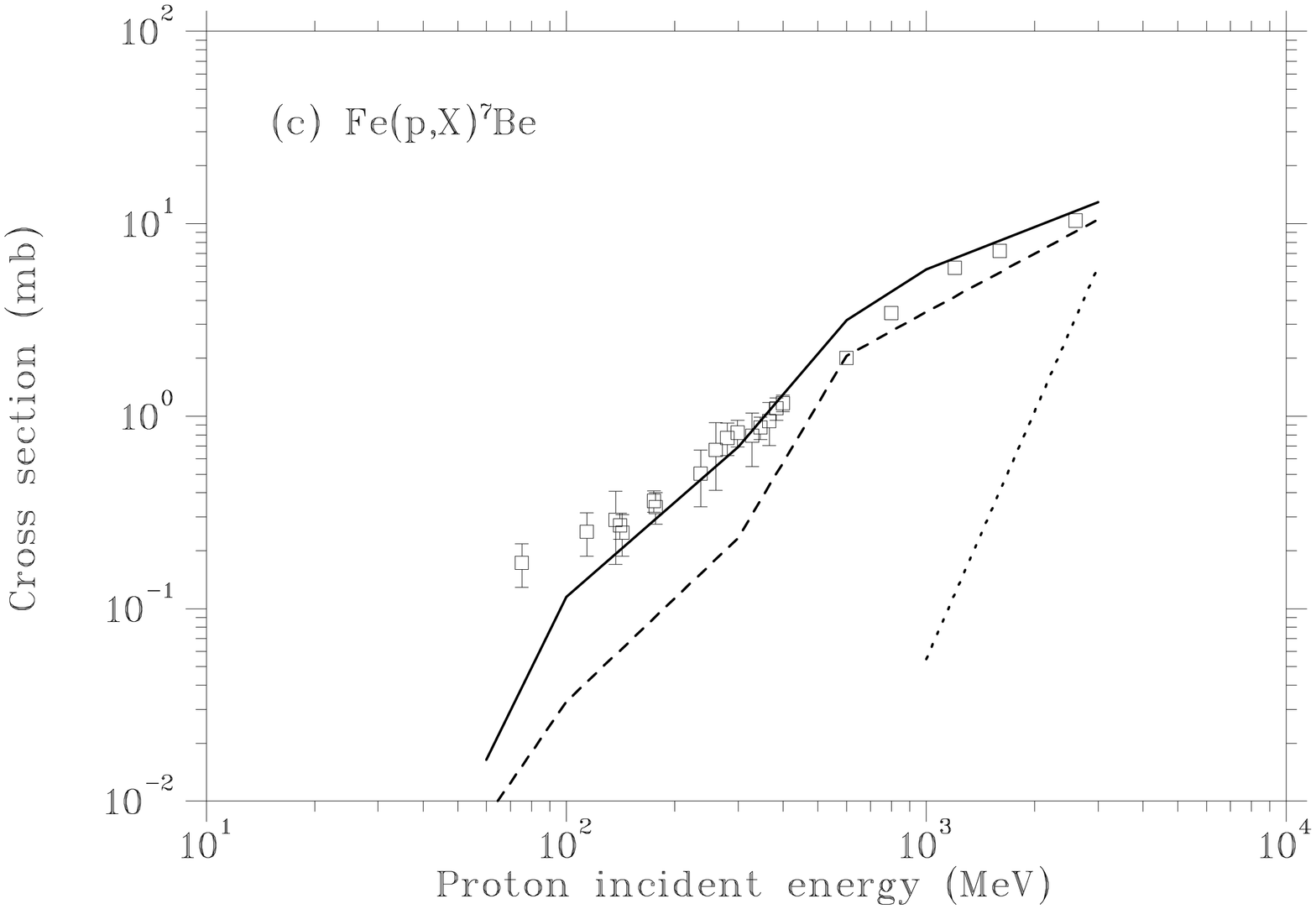}
}
&
 \rotatebox{90}{  
\includegraphics[%
         height=7cm]{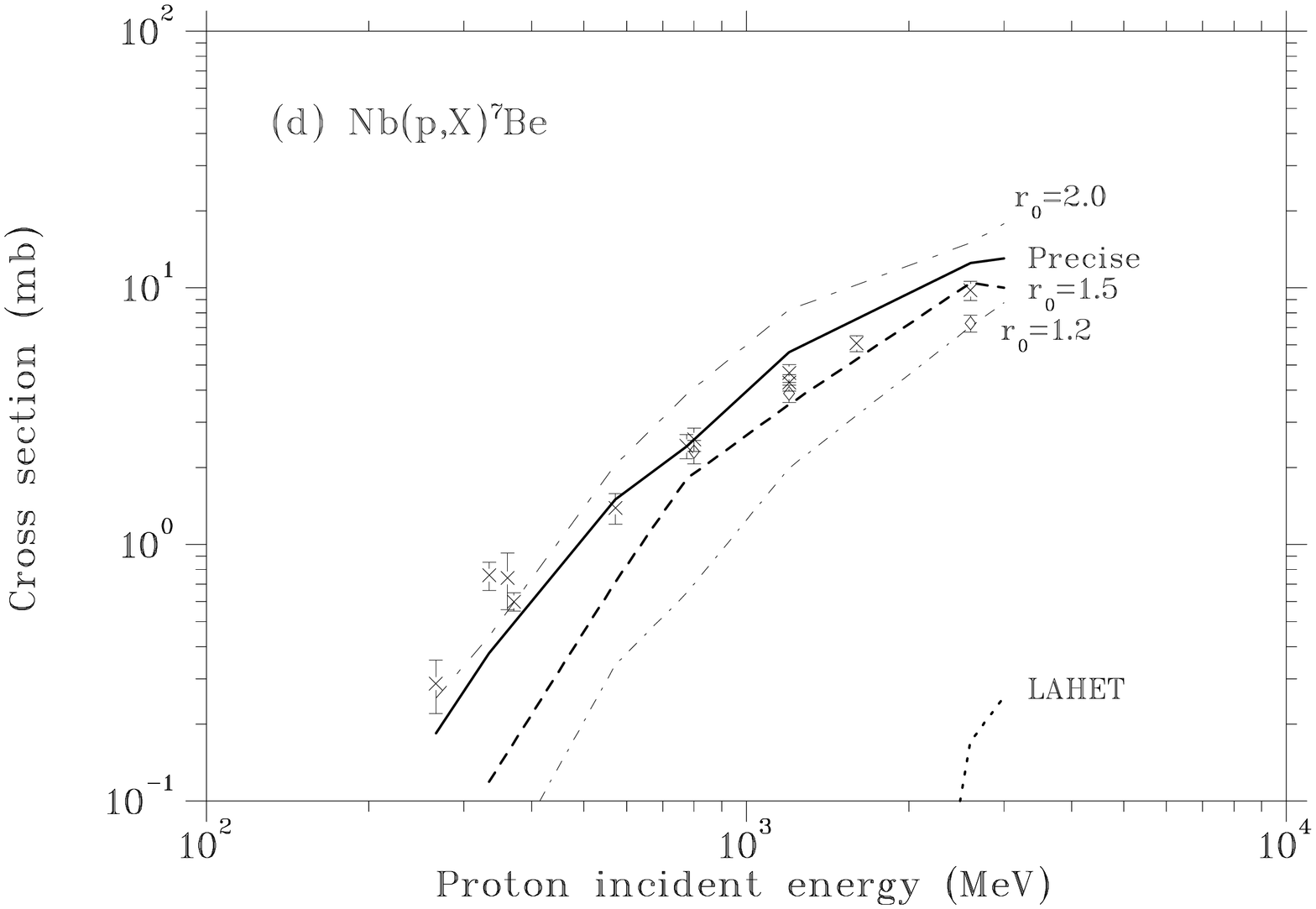}
}
\end{tabular}
\end{center}

\caption{The excitation function of $^7$Be produced from $^{16}$O,
 $^{27}$Al, $^{\rm nat}$Fe and $^{93}$Nb: 
The estimates by INC/GEVAP with $r_0=1.5$ and the precise parameter
set are shown by the dashed lines and the solid lines.
The results calculated by using LAHET are shown by the dotted lines.
The open squares are the experimental data collected in
Ref. \protect\cite{NEA}, and 
the crosses and the open diamonds are experimental data in 
Ref. \protect\cite{MIC97} and Ref. \protect\cite{DI90c}, respectively.
}\label{f-be7} 
\end{figure}

The excitation functions of $^7$Be produced by proton reactions on
$^{16}$O, $^{27}$Al, $^{\rm nat}$Fe, and $^{93}$Nb are shown
in Fig. \ref{f-be7}. 
The results calculated by INC/GEVAP with the precise parameter set,
which consists of the parameters for inverse reactions
determined by Dostrovsky {\it et al.} \cite{evap} and Matsuse {\it et
al} \cite{ref124}  and the GCCI level density parameter, are shown by
the solid lines.
The estimates by INC/GEVAP with $r_0=1.5$ and those by LAHET are also
shown in the figures as well as the experimental data collected in
Ref.\cite{NEA}. 
The results by INC/GEVAP with $r_0=1.2$ and $2.0$ are represented only
for the Nb target,  
because there is less difference in the results for other
targets. 
For the O target, the differences in the estimates between by INC/GEVAP
with $r_0=1.5$ and by that with $r_0=1.2$ or $1.5$ are 20 \%, except
at the threshold energy.  
For Al target, the differences are within a factor of two in the whole   
energy region.
INC/GEVAP with $r_0=2.0$ produces almost the same cross sections for 
the Fe(p,X)$^7$Be reaction as those with precise parameter set, and 
the differences are within 20 \%.
In the whole energy region, for all the targets, INC/GEVAP produces more
$^7$Be as $r_0$ increases.
The estimates by INC/GEVAP for Al with $r_0=1.5$ 
give the best agreement with the experimental data, as seen the
dashed line lying underneath the measurement points in the whole
energy region in Fig.\ \ref{f-be7} (b).
Whereas for Fe and Nb INC/GEVAP with the precise parameter set
reproduce the excitation functions better than that with the simple
parameter set, and the estimates agree with most of the 
experimental data within 50 \%.

INC/GEVAP reproduces the excitation functions for all the targets, 
whereas LAHET fails to reproduce the shape of the excitation functions
except for the O target.
The shapes of the excitation functions estimated by INC/GEVAP do not
change with the choice of the parameter sets.
Since LAHET severely underestimates the $^7$Be productions from Al
below 300 MeV, Fe and Nb below 3 GeV, 
it is obvious that the Fermi break-up is not the dominant
process for the $^7$Be productions in these reactions.

\begin{figure}
\begin{center}
\begin{tabular}{ll}
 \rotatebox{90}{\includegraphics[%
         height=9cm]{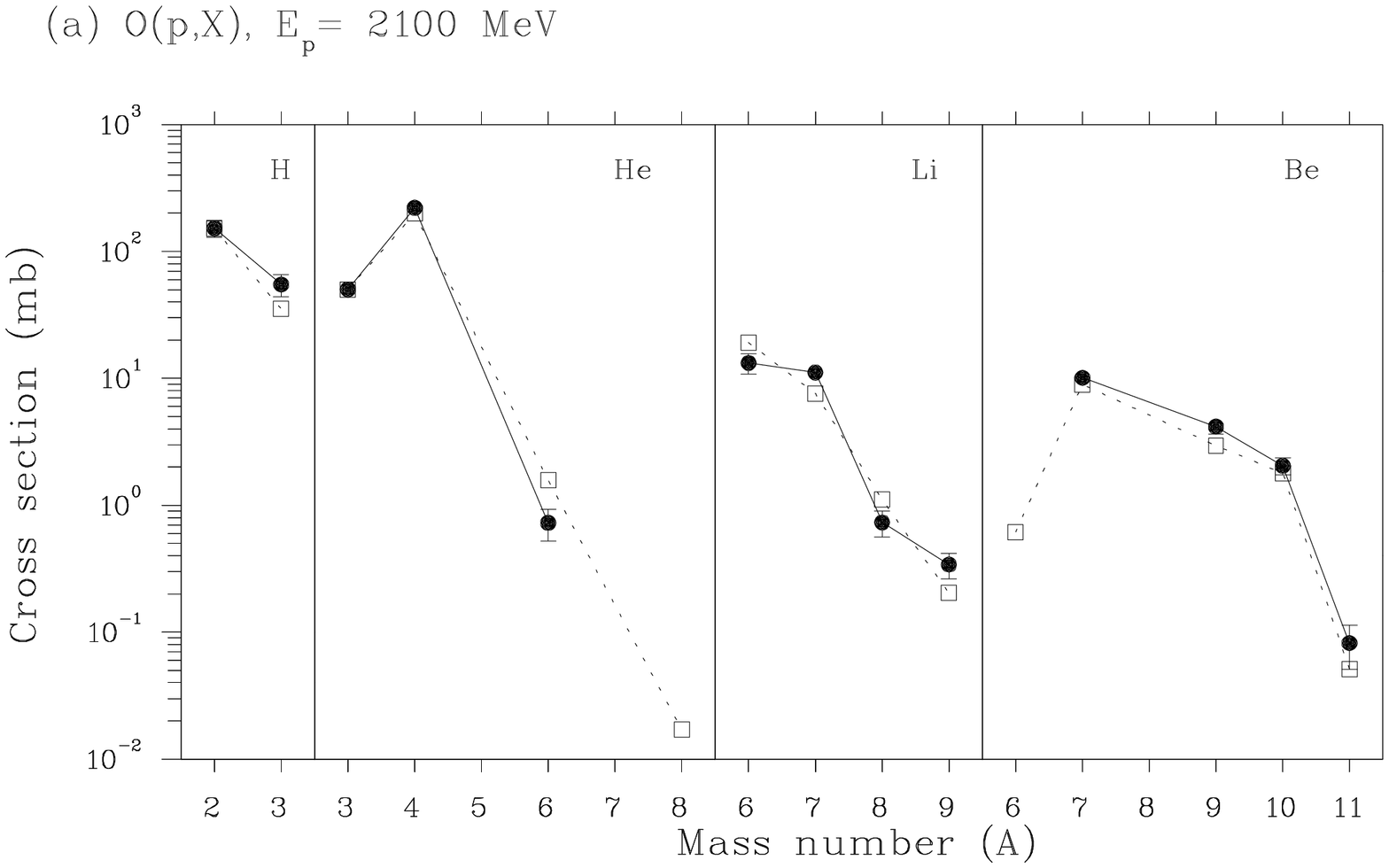} }&
 \rotatebox{90}{\includegraphics[%
         height=7cm]{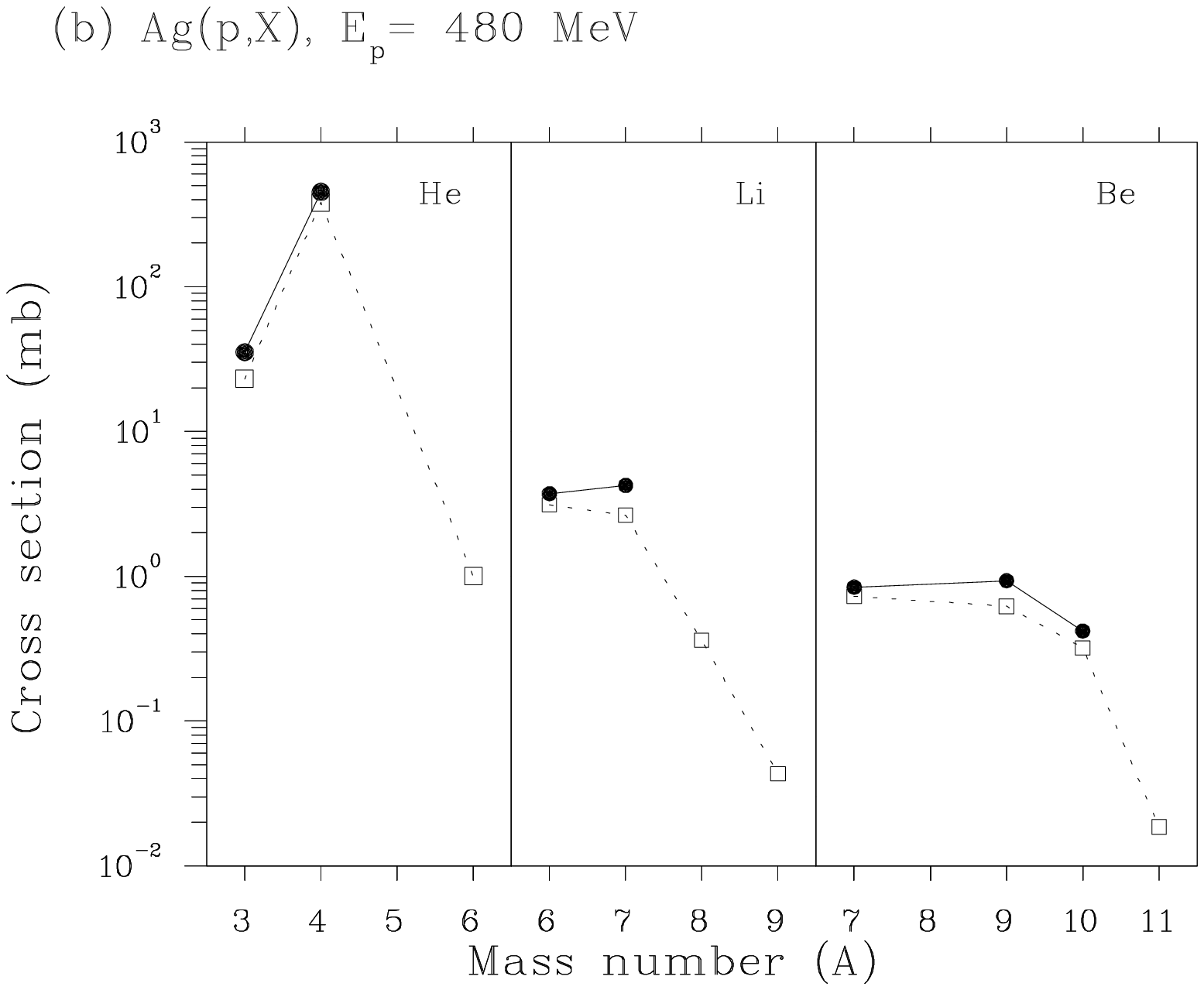} }
\end{tabular}
\end{center}

\caption{The isotopic distributions of the nuclei produced from: 
(a) 2100 MeV protons incident on $^{16}$O;
(b) 480 MeV protons incident on $^{\rm nat}$Ag.
The open squares with the solid lines denotes the results by
calculated INC/GEVAP with the precise parameter set,
and the black circles with the dashed lines denotes
experimental data \protect\cite{OLS83,ref83}.
The lines are drawn to guide to eyes.
}\label{f-iso} 
\end{figure}

The isotopic distributions of H, He, Li, and Be nuclei produced from 2100
MeV proton incident on $^{16}$O and from 480 MeV proton incident on
$^{\rm nat}$Ag are shown in Fig \ref{f-iso}. 
The results estimated by INC/GEVAP with the precise parameter set
(the open squares) are shown as well as 
the experimental data for the O target measured by
Olsen {\it et al.} \cite{OLS83}, and the data for the Ag target by Green
{\it et al.} \cite{ref83} (the closed circles).
INC/GEVAP reproduces the isotopic distributions for both these
reactions, and the estimates agree with most of the measurements with
50 \%  accuracy.

\section{Summary and Conclusion}\label{conc}

We have formulated a generalized evaporation model (GEVAP) based on the
Weisskopf-Ewing model \cite{wei2,wei1}.
The features of the model are: 
(1) the accurate level density function is used for 
deriving the decay width of particle emission;
(2) sixty-six nuclides up to Mg, not only in their ground state but also
in their excited states, are taken into account in this study.

The combination of the intranuclear cascade model (INC) and GEVAP
successfully reproduces the excitation functions of $^7$Be produced
by protons incident on $^{16}$O, $^{27}$Al,  $^{\rm nat}$Fe, and
$^{93}$Nb.
The choice of the parameter set in GEVAP does not affect the resulting 
shapes of the excitation functions.
INC/GEVAP also predicts the isotopic distributions of H, He, Li,
and Be produced from O and Ag with 50 \% accuracy.

From the results, it is concluded that the evaporation process
is the main process via which particles lighter than or equal to Be
are produced by protons incident on targets heavier than O.
INC/GEVAP can predict the production cross sections of particles
lighter or equal to Be between 50 \% to a factor of two in accuracy
depending on the choice of the parameter set used in GEVAP.
In this study, the precise parameter set gives the best results for
overall reactions, and the accuracy is 50 \% on the average.

\acknowledgements

The author would like to thank Dr. T. Numao for useful discussion and
encouragements, and also Dr. L. Moritz, Dr. G. Greeniaus,
Dr. P. Jackson and  Dr. R. Korteling for valuable comments on this
paper, and Dr. R. E. Prael for supplying the LAHET code. 
The author also appreciates the TRIUMF computer group for providing a
computer for the calculations.


\mediumtext

\begin{table}
\caption{The ejectiles considered in this study}\label{t-emittor} 
\begin{center}
{\small
\begin{tabular}{clllllll}
$Z_j$&\multicolumn{7}{c}{Ejectiles}\\\hline
0&n\\
1&p&d&t\\
2&$^{ 3}$He&$^{ 4}$He&$^{ 6}$He&$^{ 8}$He\\
3&$^{ 6}$Li&$^{ 7}$Li&$^{ 8}$Li&$^{ 9}$Li\\
4&$^{ 7}$Be&$^{ 9}$Be&$^{ 10}$Be&$^{ 11}$Be&$^{ 12}$Be\\
5&$^{ 8}$B&$^{ 10}$B&$^{ 11}$B&$^{ 12}$B&$^{ 13}$B\\
6&$^{ 10}$C&$^{ 11}$C&$^{ 12}$C&$^{ 13}$C&$^{ 14}$C&$^{ 15}$C&$^{ 16}$C\\
7&$^{ 12}$N&$^{ 13}$N&$^{ 14}$N&$^{ 15}$N&$^{ 16}$N&$^{ 17}$N\\
8&$^{ 14}$O&$^{ 15}$O&$^{ 16}$O&$^{ 17}$O&$^{ 18}$O&$^{ 19}$O&$^{ 20}$O\\
9&$^{ 17}$F&$^{ 18}$F&$^{ 19}$F&$^{ 20}$F&$^{ 21}$F\\
10&$^{18}$Ne&$^{19}$Ne&$^{20}$Ne&$^{21}$Ne&$^{22}$Ne&$^{13}$Ne&$^{24}$Ne\\
11&$^{ 21}$Na&$^{ 22}$Na&$^{ 23}$Na&$^{ 24}$Na&$^{ 25}$Na\\
12&$^{ 22}$Mg&$^{ 23}$Mg&$^{ 24}$Mg&$^{ 25}$Mg&$^{ 26}$Mg&$^{ 27}$Mg&$^{ 28}$Mg\\
\end{tabular}
}
\end{center}
\end{table}

\end{document}